\documentclass[11pt,twoside]{article}

\usepackage{asp2006}
\usepackage{epsf}
\usepackage{lscape}
\usepackage{graphicx}

\newcommand{\civ}{\ifmmode {\rm C}\,{\sc iv} \else C\,{\sc iv}\fi}
\newcommand{\nv}{\ifmmode {\rm N}\,{\sc v} \else N\,{\sc v}\fi}

\begin{document}
\title{High Velocity Outflows in Quasars}   %%% Fill in title
\author{Paola Rodr\'iguez Hidalgo\altaffilmark{1}, Fred Hamann\altaffilmark{1}, Daniel Nestor\altaffilmark{2} and Joseph Shields\altaffilmark{3}}   %%% Fill in author names
\altaffiltext{1}{Department of Astronomy, University of Florida, Gainesville, FL 32611, USA}    %%% Fill in author affiliations
\altaffiltext{2}{Institute of Astronomy, University of Cambridge, Cambridge, CB3 0HA, UK }
\altaffiltext{3}{Department of Physics and Astronomy, Ohio University, Athens, OH 45701, USA}

\begin{abstract} 

High velocity (HV) outflows are an important but poorly understood
aspect of quasar/SMBH
evolution. Outflows during the luminous accretion phase might play a
critical role in ``unveiling" young dusty AGN and regulating star
formation in the host galaxies. 
%They might even be essential to carry
%angular momentum away from the accretion disk. 
Most quasar studies have focussed on the broad absorption
lines (BALs). We are involved in a program
to study a nearly unexplored realm of quasar outflow parameter
space: HV winds with $v>10,000$ km/s up to
$v\approx 0.2c$ but small velocity dispersions (narrow absorption
lines), such that $\Delta v/v \ll 1$. Narrow-line HV flows merit
specific attention because they complement the BAL work and
pose unique challenges for
models of the wind acceleration, mass loss rates, launch radii, geometry,
etc. 
%They might also be common, e.g., accounting for a
%significant fraction of the narrow absorption lines previously attributed
%to unrelated
%(intervening) gas or galaxies. 
We have selected the brightest quasars at
$1.8<z<3.5$ with candidate narrow HV outflow lines (CIV 1548 \AA ) in existing
SDSS spectra and followed up with monitoring observations
to i) characterize, for the first time, the variability in a sample of absorbers
spanning a wide range of velocities and FWHMs, including some
marginal BALs, ii) identify/confirm more of the true outflow systems,
e.g., among the more ambiguous narrow lines,
and iii) test the
wind models and derive better constraints on basic outflow properties,
such as the kinematics, locations, and physical conditions.

\end{abstract}

%%% MAIN BODY OF TEXT GOES HERE. CONSULT "INSTRUCTIONS FOR AUTHORS USING
%%% LATEX2E MARKUP", SECTIONS 2.3-2.6 FOR HELP WITH EQUATIONS, FIGURES,
%%% AND TABLES.

%\section{Introduction}  

High velocity (HV) outflows are  fundamental constituents of active galactic
nuclei (AGN). Absorption lines that identify ejected matter are
detected in roughly 30\% to 40\% of (optically selected) quasars and
50\% to 70\% of Seyfert 1 galaxies (Crenshaw et al. 1999, Hamann \&
Sabra 2003). The outflows themselves might be present in all 
AGN if, as expected, the absorbing gas subtends only part of the sky
as seen from the central continuum source. The winds are believed
to originate in the accretion disk around the central super-massive
black hole (SMBH), driven (somehow) by accretion processes. The
need for accreting matter to expel angular momentum could mean that 
outflows are {\it essential} for SMBH growth. AGN outflows might also play a pivotal 
role in ``unveiling" young dust-enshrouded AGN, regulating star formation in the 
host galaxies (Silk \& Rees 1998, Di Matteo et al. 2005), and distributing metal-rich 
gas to the intergalactic medium, e.g., at high redshifts. 
Nonetheless, AGN outflows remain poorly understood. 

To understand the general outflow phenomenon, we need a more complete understanding of the diverse outflow types. Previous studies on quasar outflows have focused on the most obvious features, the broad absorption lines (BALs) and the narrower ``associated'' absorption lines (AALs, near the emission redshift). We have
begun a program to study a nearly unexplored part of outflow parameter
space: extremely high flow velocities, $v>10,000$ km/s (up to 0.2$c$) with low 
internal velocity dispersions, $\Delta v$ (relatively narrow absorption lines), such that 
$\Delta v/v \ll 1$, previously attributed to unrelated (intervening) gas, but that recent statistical studies confirm to be (approx. 36\%, Richards et al. 1999) AGN winds. 
These kinematic characteristics are far outside the realms of the standard classes
of AALs, with $v<5000$ km/s, and BALs, with $\Delta v\sim 10,000$ km/s and $v/\Delta v\sim 1$.  We have developed the first systematic accounting of outflow lines, taking advantage of Sloan Digital Sky Survey data (SDSS) DR4. We have analyzed the 2,200 highest signal-to-noise ratio spectra with emission redshifts between 1.8$<$z$<$3.5 to be able to observe blueshifted HV \civ\, outflows given the SDSS spectral coverage. The \civ\, candidates were fitted in order to extract quantitative information, such as v, REW and FWHM (figure \ref{fig2}).

\begin{figure}
\begin{center}
\includegraphics[scale=0.6]{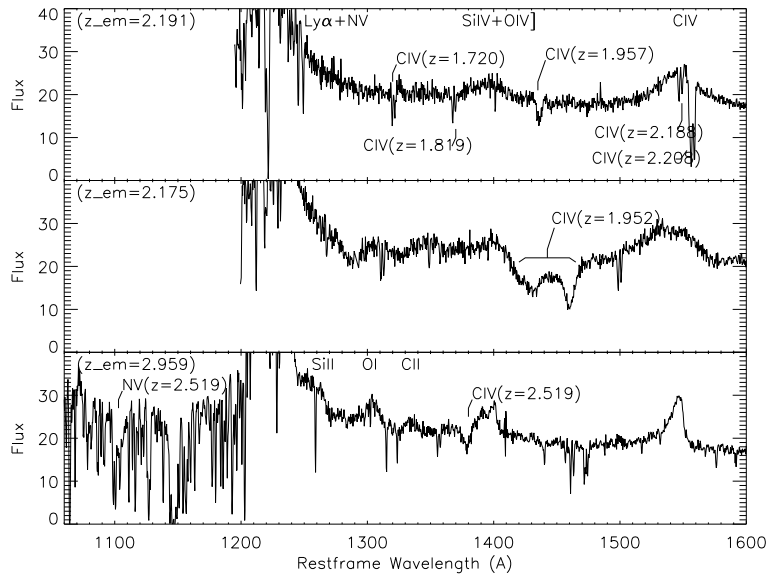}
\end{center}
\caption{Spectra of three quasars with mini-BALs, shifted to the rest wavelengths of the quasars. Prominent broad emission lines are labeled across the top in the upper and lower panels. Various candidate HV \civ\, lines are labeled with tic marks and the measured redshifts. The upper spectrum contains several \civ\, absorption systems, including two AALs, in addition to the HV mini-BAL candidates at z=1.957. The middle spectrum shows a quasar with balnicity index$>$0. The lower panel shows \nv\, absorption at the same redshift as the \civ\, mini-BAL. This spectrum also contains an unrelated damped-Ly$\alpha$ system.
\label{fig1}}
\end{figure}

%\subsection{}   %%% Second level section head (remove "%" symbol)
%\subsubsection{}   %%% Lowest level section head (remove "%" symbol)
%\section*{}    %%% Unnumbered top level section head (remove "%" symbol)
%\subsection*{}   %%% Unnumbered second level section head (remove "%" symbol)

\begin{figure}
\begin{center}
\includegraphics[width=8cm]{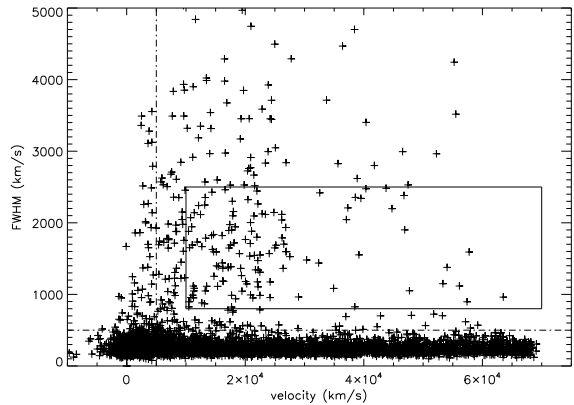}

\end{center}
\caption{The figure shows the velocity-FWHM distribution of the 5,320 \civ\,
candidates found and analyzed. Dashed-dotted lines represent selection
criteria to avoid intervening systems (FWHM$<$500 km/s) and associated
systems (v$<$10,000km/s). The plot shows continuity in the FWHM-velocity space distribution between narrow absorbers (FWHM$<$800 km/s) and broad absorbers (FWHM$>$3000 km/s). Our study focuses on the outflows confined by the square (miniBALs with velocities $>$10,000 km/s). \label{fig2}}
\end{figure}

\begin{itemize}

\item[*] Number of miniBALs found: 423
\item[*] Number of quasars with miniBALs: 284 
\item[*] Number of quasars with HV miniBALs with \\ v $>$ 10,000 km/s : 175
\item[*] Number of quasars with HV miniBALs with \\ v $>$ 25,000 km/s : 51
\end{itemize}

\begin{figure}[h!]
\begin{center}
\includegraphics[scale=0.26,angle=270]{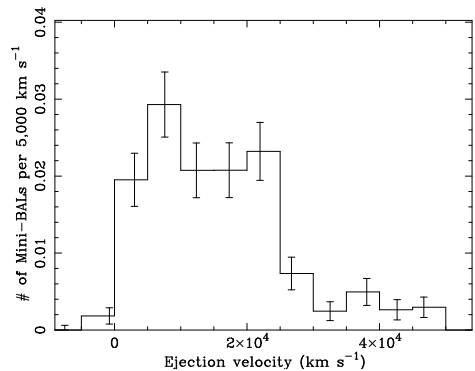}
\end{center}
\caption{Normalized number of miniBALs in 5,000 km/s ejection velocity bins\label{dndv}}

\end{figure}

Since the spectra coverage of all the quasars is not the full velocity range, not statistical results can be derived unless we "normalize" each velocity bin by the actual number of quasars with reliable measurements in that velocity range. Figure \ref{dndv} shows the normalized number of miniBALs in 5,000 km/s ejection velocity bins. For example, 2.3$\%$ of quasars have a mini-BAL between 20,000 and 25,000 km/s, and roughly 14$\%$ of quasars have a mini-BAL somewhere between 0 and 50,000 km/s (adding all the bins together).

\newpage

\begin{figure}
\begin{center}
\includegraphics[width=\columnwidth,angle=90,scale=0.6]{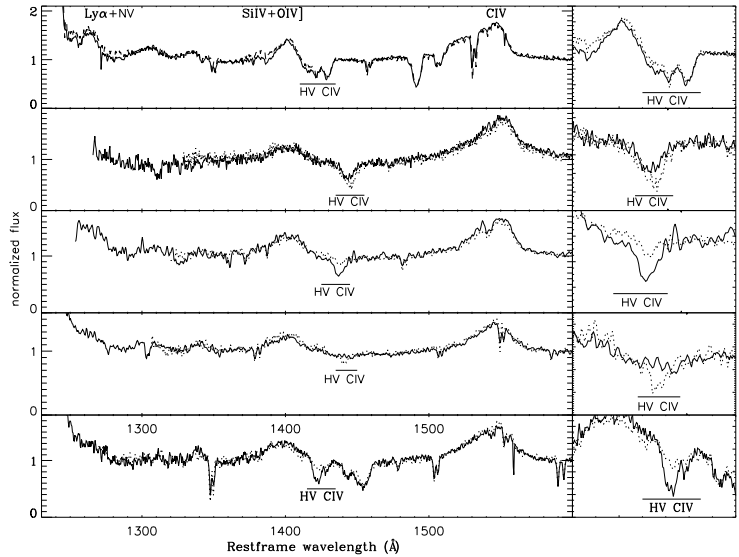}
\end{center}
\caption{Normalized spectra of the five quasars is shown shifted to their restframe wavelength. SDSS spectra is shown in dashed lines for comparison.\label{fig3}}
\end{figure}

Variability studies can confirm whether absorption is unrelated (intervening) or related to the AGN (winds) and constrain, in a systematic way, the dynamics, structure and geometry of these winds relative to their parameters (velocity and FWHM). Past works in BAL variability shows that
 half of them vary over 1-1.5 years timescales in the
 observed frame (Barlow 1993). We followed up intrinsic absorption candidates in three observing campaigns: two of them at the 2.1m telescope at the Kitt Peak National Observatory (KPNO) and one at the Michigan-Dartmouth-M.I.T observatory (MDM). Five examples of quasars where we found HV miniBALs variability, in observed frame timescales of 1-1.7 years, are presented in figure \ref{fig3}. Future work includes constraining geometrical and dynamical parameters to test theoretical models. Of special interest is the case of extreme high velocities and moderately high column densities since they can rule out
 radiative acceleration unless one invokes an uncomfortably small launch
 radius.

\acknowledgements %%% Text of acknowledgements runs on after this command.
This work is supported by the National Science Foundation via grant AST99-84040. 

%%% THE BIBLIOGRAPHY
%%%
%%% CONSULT SECTION 3 OF "INSTRUCTIONS FOR AUTHORS" FOR HOW TO USE NATBIB.
%%% AUTHORS ARE ENCOURAGED TO USE EITHER THE "THEBIBLIOGRAPY" ENVIRONMENT
%%% BY UNCOMMENTING (DELETING THE "%" SYMBOL) THE COMMANDS BELOW, OR BY
%%% USING THE BIBTEX ENVIRONMENT. TO FIND OUT WHICH IS APPLICABLE TO YOUR
%%% CONTRIBUTION, CONSULT THE VOLUME EDITORS FOR YOUR PROCEEDINGS.
%%%

\end{document}